\documentclass[a4paper,11pt]{article}
\usepackage{pos}

\usepackage{subfigure}
\usepackage{orcidlink}
\usepackage[group-separator={,},group-minimum-digits=3]{siunitx}

\title{In-situ environmental radiation background measurement in the second phase of CJPL}
\ShortTitle{In-situ environmental radiation measurement in CJPL-II}

\author*[a]{P. Zhang\orcidlink{0009-0005-0472-0130}}
\author[a]{Z. Zeng\orcidlink{0000-0003-1243-7675}}
\author[a,b]{J. Cheng}
\author[a]{H. Ma\orcidlink{0000-0001-8585-6665}}
\affiliation[a]{
  Key Laboratory of Particle and Radiation Imaging (Ministry of Education) and Department of Engineering Physics,
  Tsinghua University, Beijing 100084, China}
\affiliation[b]{School of Physics and Astronomy, Beijing Normal University, Beijing 100875,China}
\emailAdd{zhangpen21@mails.tsinghua.edu.cn}
\emailAdd{mahao@tsinghua.edu.cn}

\abstract{
China Jinping Underground Laboratory (CJPL), the deepest and largest underground laboratory worldwide, provides a low radiation background environment, which is necessary to frontier scientific research, such as the experimental studies of rare-event physics.
Due to the almost filled space of CJPL-I and the requirement of future physics experiments, the construction of the second phase of CJPL (CJPL-II) was started in 2020 and all finished in 2024.
In this work, we report the measured results of major environmental radiation in CJPL-II, including cosmic-ray muons, radon, gamma rays, and neutrons.
Results indicate that the rock overburden and the radioactive background control effectively minimize the environmental radiation background.
The scientific data presented also serve as an important basis to detector background modeling for the physics experiments in CJPL-II.
}

\FullConference{19th International Conference on Topics in Astroparticle and Underground Physics (TAUP2025)\\
24–30 Aug 2025\\
Xichang, China\\}

\begin{document}
\maketitle

\section{Introduction}
\begin{figure}[!htb]
    \centering
    \subfigure{
      \includegraphics[height=5.2cm]{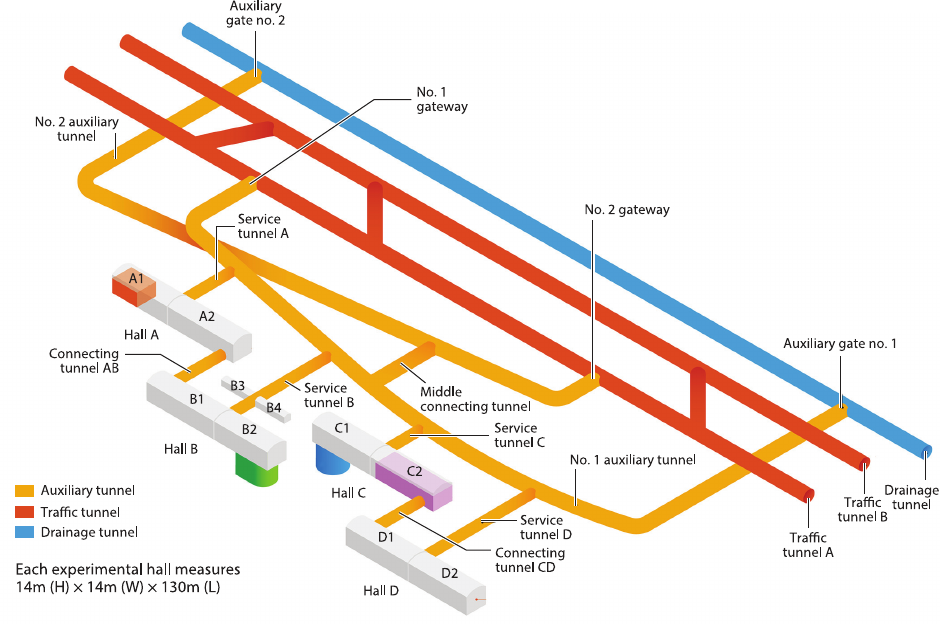}
    }
    \subfigure{
      \includegraphics[height=5.2cm]{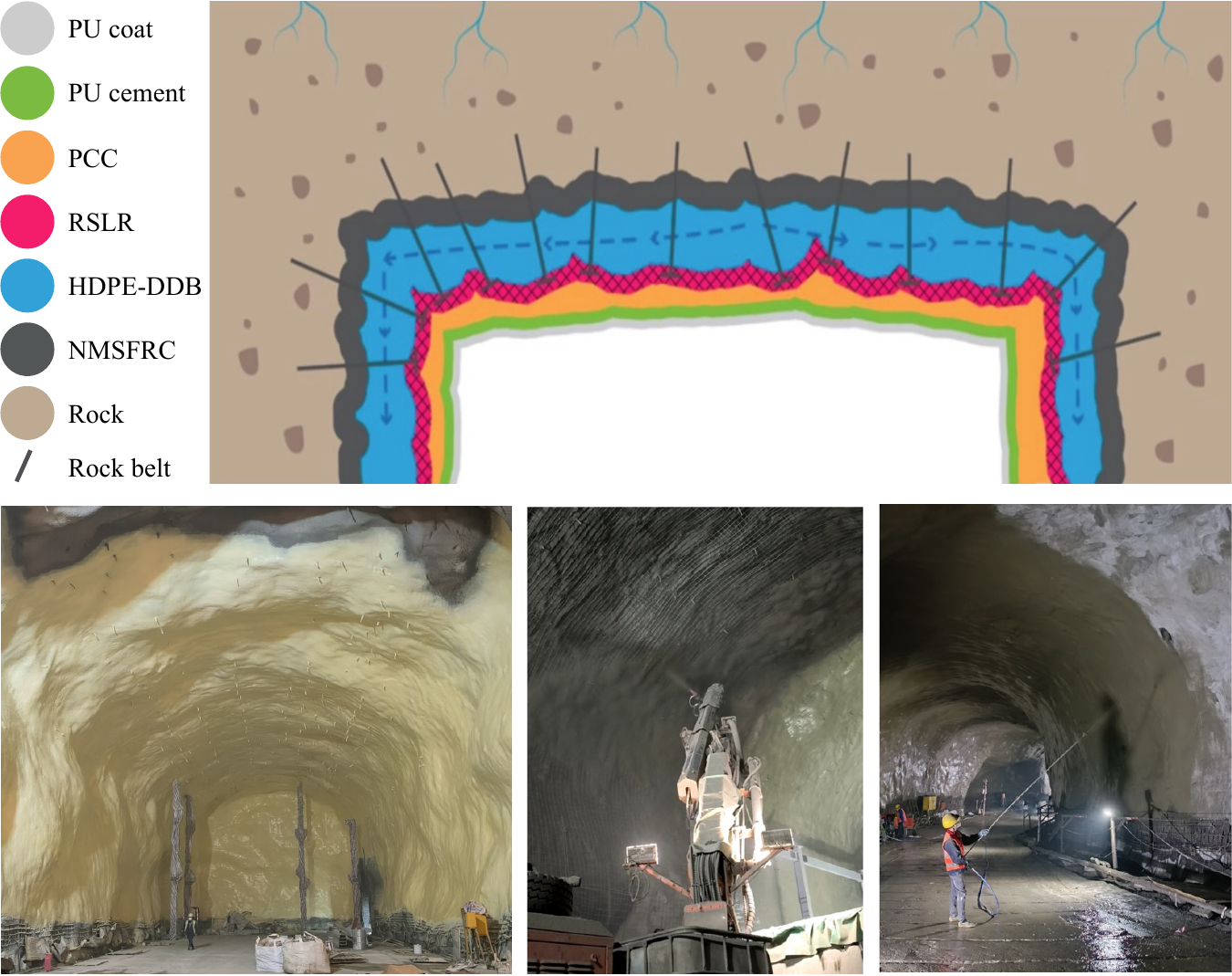}
    }
    \caption{
        Left: CJPL-II layout.
        Right: WRRS layer schematic diagram and photos taken during its construction.
        The WRRS layer was built with multiple stages applied from the rock to the surface: nano-modified steel fiber reinforced concrete (NMSFRC), HDPE dimpled drainage board (HDPE-DDB), rock bolts, rapid-Setting liquid rubber (RSLR), polymer cement concrete (PCC), polyurethane cement (PU cement), and polyurethane coat (PU coat).
    }
    \label{fig.cjpl2_wrrs}
\end{figure}

China Jinping Underground Laboratory (CJPL) is located in the middle of a traffic tunnel of the Jinping Mountain in southwest China, with a rock overburden of about 2400\,m~\cite{cheng_china_2017}.
With the ultralow radiation environment, the first phase of CJPL (CJPL-I) has attracted a great deal of interest in frontier physics research, such as dark matter direct detection and neutrinoless double beta decay searches.
Due to the almost filled space of CJPL-I and the requirement of future physics experiments, construction of the second phase of CJPL (CJPL-II), funded by National Major Science and Technology Infrastructure Construction Projects of China, was started in December 2020 and all finished in December 2024. Fig.\ref{fig.cjpl2_wrrs}-left depicts the overall layout of CJPL-II.

Various techniques, including the low-radioactivity material screening and the Water-Resistant and Radon Suppression (WRRS) layer (see Fig.\ref{fig.cjpl2_wrrs}-right), were implemented during the construction to control the radiation background in CJPL-II~\cite{xu_study_2023,zhang_measurement_2025}.
Along with the largest available laboratory space of \SI{300000}\,$\text{m}^3$ globally 
and the convenient horizontal drive-in access~\cite{cheng_china_2017},
CJPL-II is an exceptional infrastructure for next generation low-background physics experiments.
To assess the current radiation background in CJPL-II, we conducted a series of in-situ measurements of major environmental radiation, including cosmic-ray muons, radon, gamma rays, and neutrons.

\section{Radiation measurements}

\subsection{Cosmic-ray muons}
The cosmic-ray muon flux in CJPL-II is significantly reduced due to the substantial shielding provided by the overlying mountain.
To assess the residual muon flux, in 2016--2020, we deployed a plastic scintillator muon telescope system in CJPL-II and detected 161 muon events over an effective live time of 1098 days.
The detection efficiency was calculated with a two-step simulation:
a) simulating the transportation of muons through the Jinping mountain with MUSIC to obtain the energy and angular distributions of the muons reaching the underground laboratory.
b) simulating the detector response to these the underground muons in the experimental hall with Geant4.
The cosmic-ray muon flux in CJPL-II was determined to be $ (3.03 \pm 0.24 (\text{stat}) \pm 0.18 (\text{sys})) \times 10^{-10} \text{cm}^{-2} \text{s}^{-1}$,
which is the lowest among all underground laboratories worldwide~\cite{zhang_measurement_2025-1}.

\subsection{Radon and gamma rays}
In 2021--2023, with a surface radon exhalation rate detector, NRM-P01, the radon exhalation rates in different stages of the WRRS layer installation were measured to test the validity of the WRRS layer~\cite{xu_study_2023}.
After the installation was completed, the radon exhalation rate dropped below $0.1 \ \text{mBq} \cdot \text{m}^{-2} \cdot \text{s}^{-1}$,
less than 1\% of the rock exhalation rate \cite{zhang_measurement_2025}.
Furthermore, in 2024, with a radon monitor, AlphaGuard, the radon concentration in the experimental hall of CJPL-II was determined to be $(20.3 \pm 0.1)\ \text{Bq} \cdot \text{m}^{-3}$, about 10\% of the pre-construction concentration of $(208.0 \pm 0.1) \text{Bq} \cdot \text{m}^{-3}$ measured in 2017--2020~\cite{zhang_measurement_2025}.

In 2024, to evaluate the overall environmental gamma background, in-situ gamma-ray measurements were performed using a portable high-purity germanium spectrometer in several sites of CJPL-II, including the large polyethylene shielding chamber (PE chamber) in Hall-C2.
The integral count rates of the spectra in the energy range of 60--2700 keV in Hall-A1, Hall-C2 (outside of the PE chamber), and the connecting tunnel CD were in the range of 70--77\,cps.
With the measured radon concentration from AlphaGuard, the gamma spectra contributed by the nuclides in the building materials were obtained,
and the radioactivity concentrations of $^{238}$U, $^{232}$Th, and $^{40}$K in above three sites were determined to be in the range of 4.69--5.81, 5.57--6.66, and 127.00--149.44\,Bq/kg, respectively.
Last but not least, the integral count rate of the spectrum inside the PE chamber of Hall-C2 was around 2.2 cps, which was 31 times lower than the outside measurement.
\cite{zhang_environmental_2025}

\subsection{Neutron}
Neutrons are a major background for rare physics experiments.
In 2016, we measured the neutron spectrum using a Bonner multi-sphere spectrometer in CJPL-II Hall-C~\cite{hu_background_2018}.
The neutron spectrum was calculated by unfolding the raw data using the genetic algorithm, and this method was validated using a neutron source, with the neutron energy following the Maxwell-Boltzmann distribution.
The total neutron, fast neutron (1--10\,MeV), and thermal neutron (<0.5\,eV) flux were determined to be $3.7 \times 10^{-5}$, $3.8 \times 10^{-7}$, and $1.1 \times 10^{-5} \ \text{cm}^{-2} \text{s}^{-1}$, respectively.
The thermal neutron flux was validated with another independent measurement using a $^3$He tube~\cite{zeng_research_2017}.
Results indicate that the total neutron flux in CJPL-II is about 0.1\% of the ground level \cite{sato_analytical_2015}.
The neutron flux could be reduced by a further factor of \textasciitilde20--30 in the PE chamber of Hall-C2 according to experiments in CJPL-I~\cite{du_measurement_2018,zeng_thermal_2015,zeng_research_2017}.

\section{Summary}
The measured environmental radiation levels in CJPL-II were presented in this talk.
These results indicate that the overlying 2400\,m rock and various radiation background control projects during the construction of CJPL-II effectively minimize the environmental radiation background, making CJPL-II an ideal site for low-background scientific research.
The scientific data presented also serve as an important basis for detector background modeling for the physics experiments in CJPL-II.

\section{Acknowledgments}
This work was supported by
the National Natural Science Foundation of China (12425507, 12175112)
and
the National Key Research and Development Program of China (2023YFA1607101, 2022YFA1604701).
We thank CJPL and its staff for supporting this work. CJPL is jointly operated by Tsinghua University and Yalong River Hydropower Development Company.

\bibliographystyle{JHEP.bst}
\bibliography{ref.bib}

\providecommand{\href}[2]{#2}\begingroup\raggedright\begin{thebibliography}{10}

\bibitem{cheng_china_2017}
J.~Cheng, K.~Kang, J.~Li, J.~Li, Y.~Li, Q.~Yue et~al., \emph{The {China} {Jinping} {Underground} {Laboratory} and {Its} {Early} {Science}}, \href{https://doi.org/10.1146/annurev-nucl-102115-044842}{\emph{Annual Review of Nuclear and Particle Science} {\bfseries 67} (2017) 231}.

\bibitem{xu_study_2023}
P.~Xu, \emph{{Study} on the {Source} and {Prediction} of {Indoor} {Radon} {Concentration} in {Jinping} {Great} {Facility}}, {Master} thesis, Tsinghua University, 2023.

\bibitem{zhang_measurement_2025}
Q.~Zhang, P.~Xu, W.~Dai, P.~Zhang and M.~Jing, \emph{{Measurement} of radon exhalation rate from the surface of cavity and effect analysis of waterproof and radon suppression project in {China} {Jinping} {Underground} {Laboratory} phase {II}}, \href{https://doi.org/10.1360/SSPMA2024-0435}{\emph{SCIENTIA SINICA Physica, Mechanica \& Astronomica} {\bfseries 55} (2025) 111018}.

\bibitem{zhang_measurement_2025-1}
P.~Zhang, H.~Ma, W.~Dai, M.~Jing, L.~Yang, Q.~Yue et~al., \emph{Measurement of cosmic-ray muon flux at {CJPL}-{II}}, \href{https://doi.org/10.1016/j.astropartphys.2025.103147}{\emph{Astroparticle Physics} {\bfseries 172} (2025) 103147}.

\bibitem{zhang_environmental_2025}
P.~Zhang, W.~Dai, Q.~Zhang, M.~Jing, H.~Ma, Z.~Zeng et~al., \emph{{Environmental} gamma background measurement in the phase {II} of {China} {Jinping} {Underground} {Laboratory}}, \href{https://doi.org/10.1360/SSPMA-2024-0461}{\emph{SCIENTIA SINICA Physica, Mechanica \& Astronomica} {\bfseries 55} (2025) 111017}.

\bibitem{hu_background_2018}
Q.~Hu, \emph{{Background} {Research} of {Tonne}-{Scale} {Germanium} {Detector} for {Dark} {Matter} {Searches}}, {Ph}.{D}. thesis, Tsinghua University, 2018.

\bibitem{zeng_research_2017}
Z.~Zeng, \emph{{Research} and {Application} of {Low} {Background} {Thermal} {Neutron} {Detection} {Technology}}, {Ph}.{D}. thesis, Tsinghua University, 2017.

\bibitem{sato_analytical_2015}
T.~Sato, \emph{Analytical {Model} for {Estimating} {Terrestrial} {Cosmic} {Ray} {Fluxes} {Nearly} {Anytime} and {Anywhere} in the {World}: {Extension} of {PARMA}/{EXPACS}}, \href{https://doi.org/10.1371/journal.pone.0144679}{\emph{PLOS ONE} {\bfseries 10} (2015) e0144679}.

\bibitem{du_measurement_2018}
Q.~Du, S.~Lin, S.~Liu, C.~Tang, L.~Wang, W.~Wei et~al., \emph{Measurement of the fast neutron background at the {China} {Jinping} {Underground} {Laboratory}}, \href{https://doi.org/10.1016/j.nima.2018.01.098}{\emph{Nuclear Instruments and Methods in Physics Research Section A: Accelerators, Spectrometers, Detectors and Associated Equipment} {\bfseries 889} (2018) 105}.

\bibitem{zeng_thermal_2015}
Z.~Zeng, H.~Gong, Q.~Yue and J.~Li, \emph{Thermal neutron background measurement in {CJPL}}, \href{https://doi.org/10.1016/j.nima.2015.09.043}{\emph{Nuclear Instruments and Methods in Physics Research Section A: Accelerators, Spectrometers, Detectors and Associated Equipment} {\bfseries 804} (2015) 108}.

\end{thebibliography}\endgroup

\end{document}